\documentclass[prl,aps,twocolumn,showpacs,10pt]{revtex4-1}
\usepackage{graphicx}
\usepackage{amsmath}
\usepackage[table]{xcolor}

\begin{document}

\title{Enhanced Thermoelectric Performance and Anomalous Seebeck Effects in Topological Insulators}

\author{Yong \surname{Xu}$^{1,2}$}
\author{Zhongxue \surname{Gan}$^{2}$}
\author{Shou-Cheng \surname{Zhang}$^{1,2}$}
\email{sczhang@stanford.edu}

\affiliation{$^1$Department of Physics, McCullough Building, Stanford University, Stanford, California 94305-4045, USA \\
$^2$ENN Intelligent Energy Group, ENN Science Park, Langfang 065001, China}

\begin{abstract}
Improving the thermoelectric figure of merit $zT$ is one of the greatest challenges in material science. Recent discovery of topological insulators (TIs) offers new promise in this prospect. In this work, we demonstrate theoretically that $zT$ is strongly size dependent in TI, and the size parameter can be tuned to enhance $zT$ to be significantly greater than 1. Furthermore, we show that the life time of the edge states in TI is strongly energy dependent, leading to large and anomalous Seebeck effects with an opposite sign to the Hall effect. These striking properties make TIs the promising material for thermoelectrics science and technology.
\end{abstract}

%73.50.Lw	Thermoelectric effects
%72.20.Pa	Thermoelectric and thermomagnetic effects
%71.90.+q	Other topics in electronic structure (restricted to new topics in section 71)

\maketitle

The search of high-performance thermoelectric (TE) materials for efficient heat-electricity interconversion is a long-sought goal of material science~\cite{goldsmid1986,dresselhaus2007,snyder2008}. Recent discovery of topological insulators (TIs)~\cite{qi2010,hasan2010,qi2011} sheds new light on this pursuit. TIs are new quantum states of matter characterized by an insulating bulk gap and gapless edge or surface states, which are protected by the time-reversal symmetry~\cite{kane2005,bernevig2006,konig2007}. TIs share similar material properties, namely heavy elements and narrow bulk gaps, with TE materials. Consequently many currently known TIs (like Bi$_2$Te$_3$, Sb$_2$Te$_3$ and Bi$_x$Sb$_{1-x}$) are also excellent TE materials~\cite{Zhang2009,Chen2009,Xia2009,muchler2012}. The nontrivial TI edge and surface states, which were unknown in the earlier research of TE, might be advantageous in improving the thermoelectric figure of merit $zT$.

TIs are interesting for TE due to their unique electronic structure. Distinct from conventional materials, TIs support topologically protected boundary (surface or edge) states together with bulk states, and the two types of charge carriers exhibit distinct transport properties in different dimensions. However, the approximate particle-hole symmetry near the gapless Dirac point implies vanishing Seebeck coefficient, and previous works introduced a truncation~\cite{takahashi2010,murakami2011,takahashi2012} or opened a gap~\cite{ghaemi2010} in the band structure of boundary states so that a sizable Seebeck coefficient can be recovered. In contrast to all previous work, we utilize the intrinsic properties of the
boundary states, and show that the strong energy dependence of the life time naturally leads to large and anomalous Seebeck effects. Furthermore, we show that TE properties are strongly size dependent in TIs, and this size parameter, mostly ignored in previous works, can be tuned to greatly enhance $zT$. In this work, we theoretically investigated TE transport in TIs from ballistic to diffusive regions using the Landauer transport approach. We find that $zT$ of TIs changes from nearly zero to significantly larger than 1 by varying the geometric size. The finding fundamentally changes our common belief that $zT$ is an intrinsic material property. We also predict that the boundary states of TIs can have the opposite signs for the Seebeck and Hall coefficients. These striking predictions, if confirmed experimentally, can open new directions for the science and technology of thermoelectrics.

To understand TE properties of TIs, we will first discuss the definition of $zT$ that determines TE efficiency of a material. In a typical definition, $zT$ is written as
\begin{equation}
zT = \frac{\sigma S^2T}{\kappa},
\end{equation}
where $\sigma$ is the electrical conductivity, $S$ is the Seebeck coefficient, $T$ is the absolute temperature, and the thermal conductivity $\kappa$ is the sum of contributions from electrons $\kappa_e$ and lattice vibrations $\kappa_l$~\cite{goldsmid1986}. The use of this definition inexplicitly assumes that $zT$ is an intrinsic material property, independent of the geometric size. However, this basic assumption does not always hold, as we will demonstrate in TIs.

We present a general definition of $zT$ that can describe the general geometric size dependence. Using simple derivations based on thermodynamics~\cite{goldsmid2009}, $zT$ is described as
\begin{equation}
zT = \frac{G S^2 T}{K},
\end{equation}
where $G$ is the electrical conductance and $K = K_e + K_l$ is the thermal conductance. According to Ohm's scaling law in the diffusive transport regime, $G = \sigma A / L$ and Fourier's scaling law $K = \kappa A / L$, where $A$ is the cross section area and $L$ is the length of a material. The geometry factor $A/L$ cancels between $G$ and $K$. Then if $S$ is size independent, so would be $zT$. In this sense, the two definitions, Eqs. (1) and (2), are equivalent.

However, generally $zT$ can be size dependent caused by two mechanisms: (i) Ohm's scaling law and Fourier's scaling law fail; (ii) $S$ depends on the geometric size. Both mechanisms take effect in TIs. First, Ohm's scaling law does not apply to TIs, because boundary and bulk states distribute in different physical dimensions. In addition, as boundary states have mean free paths significantly longer than bulk states, it is possible to see unusual length-dependent transport behaviors, such as ballistic transport of the boundary states and diffusive transport of the bulk states. Second, the total $S$ is described as~\cite{goldsmid2009}
\begin{equation}
S = \frac{G_1 S_1 + G_2 S_2} {G_1 + G_2}.
\end{equation}
We use subscripts ``1'' and ``2'' to denote the contributions of boundary and bulk states, respectively. The size dependence of $S$ always exists in TIs. For non-diffusive transport, all individual TE quantities (including $G_1$, $G_2$, $S_1$ and $S_2$) change with increasing $L$. Even if transport becomes diffusive, varying $A$ would modify the relative contribution of boundary and bulk states, and thus change the total $S$. Therefore, we expect a strong size dependence of $zT$ in TIs. A previous work~\cite{takahashi2012}, which makes a truncation in the band structure of edge states, observed important changes in $zT$ when varying the inelastic scattering length of edge states and the cross section of the transport system.

\begin{figure}
\includegraphics[width=\linewidth]{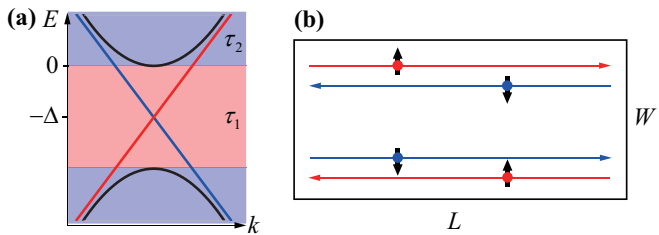}
\caption{(a) Schematic band structure and (b) schematic drawing depicting helical edge states of 2D TIs. $\tau_1$ and $\tau_2$ denote the scattering times within and outside the bulk gap, respectively. Red/blue colored lines represent edge states of upper/down spins. }
\end{figure}

We apply the Landauer transport formalism to study TE properties of TIs (see details in the Supplemental Material~\cite{SM}). We will focus on two-dimensional (2D) TIs. Similar discussion can be applied for three-dimensional (3D) TIs, which will be presented elsewhere. As a prominent feature of 2D TIs, the edge states are gapless with bands dispersing inside the bulk gap and helical with spin-momentum locking, as schematically depicted in Fig. 1. Without losing generality, we assume a linear dispersion for 1D edge states and a parabolic dispersion for 2D bulk states: ${E_{1}}(k) = \pm \hbar kv - \Delta$ and ${E_{2}}(k) = {{\hbar ^2}{k^2}}/({2m^*})$, where ``$\pm$'' denotes upper/down spins, $\hbar$ is the reduced Planck constant, $k$ is the wavevector, $v$ is the velocity of edge states, the bulk conduction band minimum (CBM) is selected as the energy reference, $-\Delta$ is the energy of the Dirac point, and $m^*$ is the effective mass.

Transport calculations usually use the constant scattering time approximation. The approximation does not rely on any assumption about the possible dependence on doping and temperature of the scattering time $\tau$, and has been successfully applied to study various TE materials~\cite{singh2010}. If assuming that the scattering rate $1/\tau$ is proportional to the density of states (DOS), like for electron-phonon scattering~\cite{goldsmid1986}, the feature that one-dimensional (1D) linear bands and 2D parabolic bands have constant DOS also suggests to use constant $\tau$.  However, care has to be taken for TIs. There are two important facts must be considered: (i) when the Fermi level $E_F$ is within the bulk gap, boundary states are protected by the time reversal symmetry against backscattering and thus have large $\tau$; (ii) when $E_F$ is outside the bulk gap, backscatterings become allowed for the boundary states due to interactions with bulk states, which decreases $\tau$ considerably.

As a generalization of the constant scattering time approximation, we introduce a dual scattering time (DST) model for TIs, as schematically depicted in Fig. 1(a). In the DST model, we assume two different constant scattering times, $\tau_1$ and $\tau_2$, for edge states with energies within and outside the bulk gap, respectively. $\tau_1$ is much greater than $\tau_2$.  Meanwhile, we assume a constant scattering time of $\tau_2'$ for bulk states. $\tau_2'$, similar as $\tau_2$, is significantly smaller than $\tau_1$. For simplicity we take $\tau_2' = \tau_2$. Note the scattering-time ratio $r_{\tau} = \tau_1/\tau_2$ is system dependent. In the HgTe quantum well, a well known 2D TI system~\cite{bernevig2006,konig2007}, $r_{\tau}$ is on the order of $10^3$ as deduced from existing experiments~\cite{daumer2003, gusev2011}. In principle $r_{\tau}$ can be enhanced, for instance, by introducing nonmagnetic defects or disorders into the system. Refs.~\cite{takahashi2010,murakami2011,takahashi2012} truncate the band structure of edge states, which can be mathematically viewed as the limiting case of $r_{\tau} = \infty$.

\begin{figure}
\includegraphics[width=\linewidth]{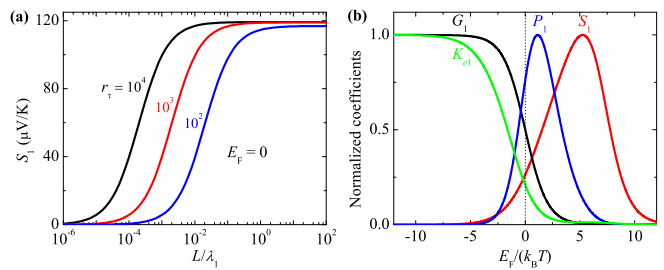}
\caption{(a) The length dependence of $S_1$ ($S$ of edge states) for $E_F = 0$ and varying scattering-time ratio $r_{\tau} = \tau_1/\tau_2$ (see $\tau_1$ and $\tau_2$ in Fig. 1). (b) The Fermi level dependence of TE quantities of edge states for diffusive transport and $r_{\tau} = 10^{3}$. $S_1$, $G_1$,  $P_1$ and $K_{e1}$ are normalized by 450 $\mu$V/K, $(2e^2/h)(\lambda_1 /L)$, $(2.5k_B^2/h) (\lambda_1 /L)$, and $(6.6 k_B^2 T /h) (\lambda_1 /L)$, respectively, where $\lambda_1$ is the inelastic mean free path of edge states and $L$ is the transport length. }
\end{figure}

Let us first focus on edge states. Herein $L$ is scaled by the inelastic mean free path of edge states $\lambda_1 = v \tau_1$, $E_F$ is scaled by $k_B T$, $\Delta$ is unimportant to calculations, and $r_{\tau}$ is the only remaining parameter. The length dependence of $S_1$ ($S$ of edge states) is visualized in Fig. 2(a) for a specified Fermi level of $E_F = 0$ and varying $r_{\tau}$. $S_1$ is nearly zero in the ballistic limit (small $L$), gradually increases with $L$, and finally becomes saturated in the diffusive limit ($L$ on the order of $\lambda_1$). This basic trend is independent of the selection of $r_{\tau}$, which only affects the results quantitatively.  Increasing $r_{\tau}$ from $10^2$ to $10^4$ leads to a larger slope in the $S_1$-$L$ curve and similar diffusive $S_1$ of about 120 $\mu$V/K. The results are consistent with previous ones~\cite{takahashi2010,murakami2011,takahashi2012} which were calculated for the limiting case of $r_{\tau} = \infty$ and $L = \lambda_1$. The optimization of $E_F$ further enhances $S_1$. As presented in Fig. 2(b) for $r_{\tau} = 10^3$, the diffusive $S_1$ can be enhanced to 450 $\mu$V/K when choosing $E_F \sim 5 k_BT$. As $E_F$ shifts upwards across the bulk CBM, $G_1$ and $K_{e1}$ sharply decrease, and the power factor $P_1 = G_1 S_1^2$ shows a peaked shape centered at $E_F \sim 1 k_B T$. At the optimal $E_F$ of $P_1$, $S_1$ is about 200 $\mu$V/K , which is comparable to the best $S$ of around 250 $\mu$V/K in bulk Bi$_2$Te$_3$ systems~\cite{goldsmid1986}.

At this point, it is important to emphasize novel aspects of the Seebeck effects of edge states. $S$ is usually small for gapless band structures, since both electron and holes are thermally excited and contribute opposite Seebeck coefficients which cancel with each other.
This explains the vanishingly small $S_1$ in the ballistic limit. However, the extraordinarily large $S_1$ obtained in the diffusive limit seems illusive. Furthermore, there exists an anomaly in the sign of $S_1$.  It is well known that the type of charge carriers, \emph{n} or \emph{p}, is defined by the sign of $S$ or the Hall coefficient. As far as we know, previous work always found the same sign in the two coefficients at least for a single type of charge carriers. In contrast, we show a counter example here. For $E_F$ around the bulk CBM, edge states with energies above the Dirac point are mostly unoccupied [see band structures of realistic systems~\cite{qi2011,xu2013}]. The corresponding charge carriers are obviously \emph{n}-type in the sense of Hall measurement, yet they contribute \emph{p}-type Seebeck effect, as evidenced by the positive sign of $S_1$.

To understand the anomalous sign of the Seebeck effects, we first explain generally how the sign of $S$ is defined. The Seebeck effect represents the response of electrons to the external temperature gradient. When increasing the temperature, the occupation of electrons gets larger or smaller, depending on whether energy of electrons is above or below ${E_{\rm{F}}}$. Electrons above or below ${E_{\rm{F}}}$ thus have opposite contribution to $S$. The Landauer formula of $S$ brings in contribution of states within around $5k_{\rm{B}} T$ of ${E_{\rm{F}}}$~\cite{singh2010}. When electrons above $E_{F}$ have dominating contribution to the Seebeck effect, the sign of $S$ is negative, and vice versa. Now let us consider the case of edge states for $E_F = 0$ (referenced to the bulk CBM). The ballistic transmission of edge states is energy independent~\cite{SM}. Electrons below and above $E_{F}$ have the same contribution to $S$, leading to a zero ballistic $S_1$. As $L$ increases, electrons above the $E_{F}$ experience more scatterings than those below the $E_F$ due to the edge-bulk interactions. As a result, $S_1$ becomes nonzero and positive. We thus conclude that the anomalous sign in $S_1$ originates from the unique energy dependence of scattering time in TIs. An anomalous sign in $S_1$ may also appear when $E_F$ is around the bulk valence band maximum (VBM).

A simple way to estimate $S$ is to use the Sommerfield expansion~\cite{paulsson2003}
\begin{equation}
 S = {\left. { - \frac{{{\pi ^2}k_{\rm{B}}^2T}}{{3e}}\frac{{\partial \ln [\overline {\cal T} (E)]}}{{\partial E}}} \right|_{E = {E_{\rm{F}}}}},
\end{equation}
which is obtained from the Landauer formula by assuming a smooth transmission function $\overline {\cal T} (E)$ and low $T$. $\overline {\cal T} (E)$ is determined by the distribution of conduction modes $M(E)$ and the mean free path  $\lambda(E)$~\cite{SM}. The formula states that $S$ is determined by the slope of the transmission function at ${E_{\rm{F}}}$, which suggests two mechanisms to enhance $S$: (i) an increased energy dependence of $M(E)$, for instance by a local increase in the density of states~\cite{mahan1996}; (ii) an increased energy dependence of $\lambda(E)$.  The later mechanism, usually thought to be unimportant, is ignored in most theoretical studies. However, it plays a crucial role in TIs. The strong energy dependence of $\lambda(E)$ (or $\tau(E)$), caused by edge-bulk interactions, makes a dominating contribution to $S_1$. This explains the large magnitude of $S_1$ obtained from the gapless band structure of the boundary states. Previous study~\cite{ghaemi2010} suggests to tune a hybridization gap in boundary states for increasing $S$ with decreasing mobility as a sacrifice. In contrast we show that edge states can contribute large $S$ by optimizing $E_F$ with no need of opening the band gap. Thus edge states can simultaneously have large $S$ and superior mobility, advantageous for TE applications.

The improvement of $zT$ requires suppressing thermal conduction while keeping electrical conduction less affected. Edge states are quite promising for this purpose, since their low physical dimension and excellent transport ability enable an effective decoupling between electrons and phonons. Specifically, as the width of 2D TIs decreases, transport of edge states remains unchanged but the lattice thermal conductance lowers; when non-magnetic perturbations (e.g. defects or disorders) are introduced into the transport system, edge states are topologically protected against scattering while phonons are significantly scattered.

What is the maximum possible $zT$ ($zT_{\rm{max}}$) a 2D TI can have? To answer this question, we consider the limiting case that edge states contribute all electrical conductance and lattice thermal conductance is negligible, which gives the best $zT$. Edge states are treated in the diffusive transport region for improving $S$ and  $zT$. Then the optimization of Fermi level gives $zT_{\rm{max}}$ as a function of $r_{\tau}$. As shown in Fig. 3(a), $zT_{\rm{max}}$ increases monotonically with increasing $r_{\tau}$. When varying $r_{\tau}$ from $10^2$ to $10^3$, $zT_{\rm{max}}$ enhances from 4 to 11, shifting the optimized $E_F$ from $2.0 k_B T$ to $3.3 k_B T$. Importantly, $zT_{\rm{max}}$ is temperature independent and much larger than 1.

\begin{figure}
\includegraphics[width=\linewidth]{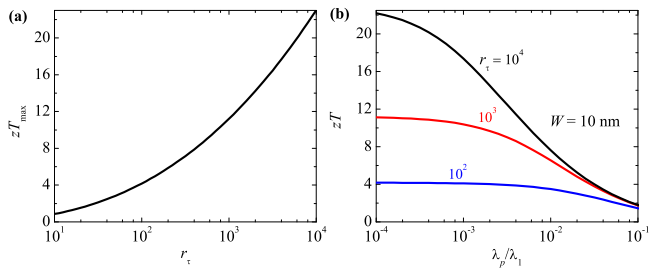}
\caption{(a) The maximum possible $zT$ of 2D TI, $zT_{\rm{max}}$, as a function of the scattering-time ratio $r_{\tau}$. (b) $zT$ including contributions of edge states and phonons as a function of $\lambda_p/\lambda_1$ for varying $r_{\tau}$, where $\lambda_p$ is the phonon mean free path and $\lambda_1$ is the inelastic mean free path of edge states. The width of transport system $W$ is selected to be 10 nm.}
\end{figure}

It is interesting to see how $zT_{\rm{max}}$ changes when including contribution of lattice thermal conduction. In our discussion, the width of transport system ($W$) is selected to be on the order of two times the localization width $\xi$ of edge states to minimize $K_l$ and to avoid hybridization between edge states~\cite{SM}. Note that inelastic scatterings of edge states can happen at finite temperatures, and the temperature dependence of $\lambda_1$ could be important. In TE calculations, the relevant quantities are the ratios $r_{\tau}$ and $\lambda_p / \lambda_1$ ($\lambda_p$ is the phonon mean free path). The ratios can vary with temperature for a given transport system. They can also be tuned at a fixed temperature, for instance by the control over disorders. To account for these effects, we calculate $zT$ for different ratio values. As presented in Fig. 3(b), an increased $\lambda_p / \lambda_1$  leads to a decreased $zT$, and such a trend looks more obvious for larger $r_{\tau}$. For instance, when changing $\lambda_p / \lambda_1$ from $10^{-3}$ to $10^{-2}$, $zT$ decreases from 4.1 to 3.5 for $r_{\tau} = 10^2$ and from 10.4 to 6.5 for $r_{\tau} = 10^3$. As discussed above, a small $\lambda_p / \lambda_1$ is in principle feasible in 2D TIs, for instance, by using non-magnetic defects or disorders. Our results thus indicate that $zT$ can keep much larger than 1 when $K_l$ is included.

Finally we take electrical conduction of bulk states into account. We perform an example study on a realistic 2D TI material, fluorinated stanene as described in the Supplemental Material~\cite{SM}, which has an nontrivial bulk gap of 0.3 eV, suitable for room temperature operation~\cite{xu2013}. Figure 4 presents the size dependence of $zT$ and $S$, whose values are maximized by optimizing $E_F$. For small $L$ and large $W$, bulk states dominate TE transport, resulting in small $zT$ and negative $S$. When increasing $L$ and decreasing $W$, edge states becomes increasingly important, leading to a bulk-edge crossover. Consequently, $zT$ improves noticeably, and $S$ has a sign change from positive to negative. The contribution of edge states is maximized by choosing $L$ on the order of $\lambda_1$ and $W$ on the order of $2\xi$ ($\sim 10$ nm). At this optimized geometry, a maximal $zT$ of 7 is realized. In practice, $\tau(E)$ may not change sharply from $\tau_1$ to $\tau_2$ as described by the DST model. We considered a smooth decrease of $\tau(E)$ from $\tau_1$ to $\tau_2$ in the form of $\exp [E/(k_B T)]$. The predicted $zT$ slightly decreases but still keeps to be extremely large (about 6).

\begin{figure}
\includegraphics[width=\linewidth]{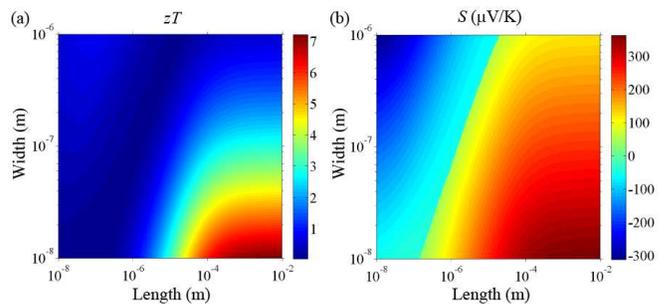}
\caption{The size dependence of (a) $zT$ and (b) $S$ for the 2D TI fluorinated stanene at 300 K.}
\end{figure}

It has been theoretically predicted~\cite{hicks1993-2D,hicks1993-1D} and then experimentally confirmed~\cite{venkatasubramanian2001,harman2002,poudel2008,hochbaum2008,boukai2008} that low-dimensional and nanostructured materials can have $zT$ much larger than their bulk counterparts. However, in those previous works a subtle control of material composition and structure is required to get an overall balance between electrical conduction and thermal conduction for optimizing $zT$. Here we propose to use TIs for TE, for which the optimization of the geometric size can suppress thermal conduction while keep electrical conduction little affected. In comparison, our approach of improving $zT$ is simpler and more effective, which could greatly prompt the development of TE science and technology.

In summary, we present the basic design principles to optimize $zT$ for TI materials. We show that $zT$ is no longer an intrinsic material property, but strongly depends on the geometric size in TIs. This new tuning parameter can dramatically increase $zT$ of topological materials, including quantum anomalous Hall insulators and topological crystal insulators. In 2D TIs, we show that $zT$ could be improved to be significantly larger than 1 by optimizing the geometric size. Moreover, we predict that the gapless edge states can contribute large and anomalous Seebeck effects with an opposite sign to the Hall effect. This striking prediction can be used to experimentally test the theoretical framework presented in this work.

\begin{acknowledgements}
We thank Yayu Wang, Biao Lian, Jing Wang, Hai-Jun Zhang and Xiaobin Chen for helpful discussions.
\end{acknowledgements}

%\bibliography{ZT}
%Merlin.mbs v4.21 2009-07-09.
%
\end{document}